\documentclass[%
 reprint,
superscriptaddress,
preprintnumbers,
nofootinbib,
 amsmath,amssymb,
 aps,
 prl,
longbibliography,
twocolumn 
]{revtex4-1}  

\usepackage[pdftex]{graphicx}
\usepackage[utf8]{inputenc}
\usepackage{dcolumn}
\usepackage{bm}
\usepackage{amsmath}

\usepackage[colorlinks,allcolors=blue,linktocpage]{hyperref}
\usepackage{cleveref}
\crefname{equation}{Eq.}{Eqs.}%

\newcommand*\diff{\mathop{}\!\mathrm{d}}
\newcommand*\Diff[1]{\mathop{}\!\mathrm{d^#1}}
\renewcommand\L{\mathcal{L}}

\begin{document}

\title{Einstein-Cartan Portal to Dark Matter}
\author{Mikhail Shaposhnikov}
\author{Andrey Shkerin}
\author{Inar Timiryasov}
\author{Sebastian Zell}
\affiliation{Institute of Physics, Laboratory for Particle Physics and Cosmology,
\'{E}cole Polytechnique F\'{e}d\'{e}rale de Lausanne, CH-1015 Lausanne, 
Switzerland}

\begin{abstract}
It is well-known since the works of Utiyama and Kibble that the gravitational force can be obtained by gauging the Poincar{\'e} group, which puts gravity on the same footing as the Standard Model fields. The resulting theory -- Einstein-Cartan gravity -- inevitably contains four-fermion and scalar-fermion interactions that originate from torsion associated with spin degrees of freedom. We show that these interactions lead to a novel  mechanism for producing singlet fermions in the Early Universe. These fermions can play the role of dark matter particles. The mechanism is operative in a large range of dark matter particle masses: from a few keV up to $\sim 10^8~$GeV. We discuss potential observational consequences of keV-scale dark matter produced this way, in particular for right-handed neutrinos. We conclude that a determination of the primordial dark matter momentum distribution might be able to shed light on the gravity-induced fermionic interactions.
\end{abstract}

\maketitle

\emph{Introduction.}---Gravity is a universal force, and it inevitably reveals some properties of yet-to-be-discovered constituents of Nature -- dark matter (DM) and dark energy (see, e.g., \cite{Feng:2010gw, Bertone:2016nfn, Arun:2017uaw} for reviews). Indeed, all evidence for DM in galaxies, galaxy clusters and at large scales is due to the gravitational interaction. Thus, we know well about how DM gravitates, but little about what it is. The list of DM particle candidates includes, e.g., weakly interacting massive particles (see \cite{Pagels:1981ke, Weinberg:1982tp, Goldberg:1983nd, Ellis:1983wd, Ellis:1983ew} for important early contributions), sterile neutrinos \cite{Dodelson:1993je}, axions \cite{Preskill:1982cy, Abbott:1982af, Dine:1982ah} and axion-like particles \cite{Hu:2000ke}. Regardless the nature of DM, any model of it should explain how it was produced in the Early Universe and how its abundance is maintained.

In this paper, we argue that gravity might be able to tell us not only about the distribution of DM in the Universe, but also about the mechanism of its production. The two key ingredients of the framework allowing for such a statement
are the Einstein-Cartan (EC) formulation of gravity \cite{Utiyama:1956sy, Kibble:1961ba} and the assumption about the fermion nature of a DM candidate.
 Regarding the second one, we take a DM particle to be a Dirac or Majorana fermion which is a singlet under the gauge group of the SM.

We begin with a brief discussion of EC gravity.\footnote{See, e.g., \cite{Hehl:1976kj, Shapiro:2001rz} for reviews.} There is no doubt that below the Planck scale, General Relativity (GR) provides an elegant and accurate description of gravity. Nevertheless, this leaves still unanswered the question about which formulation of GR one should employ. An important alternative to the most widely used metric gravity is the EC formulation. In this theory, the role of fundamental fields is played by the tetrad and the spin connection, in terms of which the metric and the Christoffel symbols are introduced. The latter are, in general, not symmetric in the lower indices, hence EC gravity contains torsion. Still, the number of propagating degrees of freedom -- two of the graviton -- is the same as in the metric formulation. A conceptual advantage of EC gravity is that it can be viewed as a gauge theory of the Poincar{\'e} group, thus allowing for a similar treatment of all fundamental forces.\footnote{Moreover, we note that in EC gravity there is no need for the Gibbons-Hawking-York boundary term \cite{York:1972sj, Gibbons:1976ue} for the variational problem to be well-posed.}

In the absence of matter, the EC and metric formulations of gravity are equivalent. This changes once scalar fields coupled non-minimally to the Ricci scalar are introduced. The resulting theory is then equivalent to the Palatini formulation of gravity \cite{Palatini1919, Einstein1925} (see also \cite{Ferraris1981}). In this version of gravity, the metric and the Christoffel symbols are viewed as independent fundamental variables, and the latter are assumed to be symmetric in the lower indices. Another way to break the equivalence of the EC and metric formulations is to take into account fermions since the latter source torsion. The theory admits an exact solution for the torsionful part of the connection. Plugging this solution back in the action, one arrives at an equivalent theory in the metric formulation. The difference of the two theories of gravity is then manifested in the appearance of dimension-six terms representing an interaction of fermionic axial currents \cite{Kibble:1961ba, osti_4843429}.\footnote{Of course, one could have started from the beginning in a metric theory with additional higher-dimensional operators. In this case, however, a consistent effective field theory approach would dictate that all possible higher-dimensional interactions (consistent with relevant symmetries) are taken into account.} Their strength is fixed and suppressed by $1/M_P^2$.

In the statements thus far, we have considered in EC gravity the same action as in the metric theory. Due to the presence of torsion, however, one can form more terms of mass dimension not bigger than four than in the metric case. Specifically, the fermion kinetic terms can be generalized by introducing non-minimal fermion couplings \cite{hep-th/0507253, Alexandrov:2008iy, 1104.2432, 1212.0585}. In an equivalent metric theory, they lead to vector-vector, axial-vector and axial-axial fermion current interactions. Furthermore, the gravitational action of the EC theory can be extended by adding the Holst term \cite{Hojman:1980kv, Nelson:1980ph, Castellani:1991et, Holst:1995pc}, which modifies the four-fermion interaction. Note that the additional terms come with \textit{a priori} unknown coupling constants and the strength of current-current interactions depends on these couplings.

The present paper uses the results of \cite{Shaposhnikov:2020frq}, where a theory of scalar and fermion fields coupled to gravity was studied. There, we included additional terms of mass dimension not bigger than four that are specific to the EC formulation and derived the equivalent metric theory. When applied to cosmology and experiment, the scalar field can be associated with the SM Higgs field, and the fermions with the SM quarks and leptons as well as, possibly, additional species such as right-handed neutrinos. The phenomenology of the scalar-gravity part of the EC theory has already been investigated in \cite{Shaposhnikov:2020geh}. There, we considered the Higgs field as an inflaton. Our study was motivated by the well-known fact that the Higgs field can be responsible for inflation provided that it couples non-minimally to the Ricci scalar \cite{Bezrukov:2007ep}. The models of Higgs inflation in the metric \cite{Bezrukov:2007ep} and Palatini \cite{Bauer:2008zj} formulations of gravity find their natural generalization within the EC theory \cite{Shaposhnikov:2020frq} (see also \cite{Langvik:2020nrs}).

The goal of this paper is to study phenomenology of the fermionic sector of the EC theory. Namely, we show that the four-fermion interaction originating from EC gravity can be responsible for DM production. We compute the abundance and the spectrum of produced (Dirac or Majorana) particles and show that the right amount of DM can be generated for a wide range of fermion masses. We also discuss an interesting case of warm DM where the primordial  momentum distribution characteristic for EC gravity can potentially be observable. \\

\emph{Einstein-Cartan gravity and fermions.}---In this work, we focus on the fermion-gravity part of the general theory studied in \cite{Shaposhnikov:2020frq}. To simplify the presentation, we only keep the Einstein-Hilbert term and the non-minimal couplings of fermions in the action.  We comment on the inclusion of other terms later; see also appendix A.
Then, for each (Dirac or Majorana) fermion species $\Psi$ the relevant part of the action reads as follows \cite{hep-th/0507253, Alexandrov:2008iy, 1104.2432, 1212.0585}:\footnote{We work in natural units $\hbar=c=1$ and use the metric signature $(-1,+1,+1,+1)$. The matrix $\gamma^5$ is defined as $\gamma^5=-i\gamma^0\gamma^1\gamma^2\gamma^3$.}
\begin{equation}\label{S_f}
\begin{split}
\L & =\frac{1}{2}M_P^2R + \frac{i}{2} \bar{\Psi}(1-i\alpha -i\beta\gamma^5)\gamma^\mu D_\mu\Psi \\  & -\frac{i}{2}\overline{D_\mu\Psi}(1+i\alpha+i\beta\gamma^5)\gamma^\mu\Psi\;,
\end{split}
\end{equation}
where $M_P = 2.435\times 10^{18}$~GeV is the Planck mass and $D_\mu$~is the covariant derivative containing the connection field. The real couplings $\alpha$, $\beta$ are chosen to be the same for all generations of fermions which implies the universality of gravity in the fermionic sector. Allowing for the couplings to depend on a generation index yields qualitatively the same results. In metric gravity, the non-minimal terms sum up to a total derivative, but in the torsionful case they contribute to the dynamics of the theory. 

The theory (\ref{S_f}) can be resolved for torsion explicitly \cite{Kibble:1961ba, osti_4843429,hep-th/0507253, Alexandrov:2008iy, 1104.2432, 1212.0585,Shaposhnikov:2020frq}. Upon substituting the solution for torsion back to the action, one obtains an equivalent metric theory with extra higher-dimensional fermion interaction terms. They read:
\begin{equation}\label{Lint}
\scalebox{0.92}[1]{
$\L_{4f}= \dfrac{3\alpha^2}{16 \,M_P^2} V^\mu V_\mu + \dfrac{3 \,\alpha \,\beta}{8\,M_P^2} V^\mu A_\mu
-\dfrac{3-3\beta^2}{16\,M_P^2} A^\mu A_\mu $}\;,
\end{equation}
where $V^{\mu} =\bar{N} \gamma^{\mu} N+\sum_{X} \bar{X} \gamma^{\mu} X$ is the vector fermion current and $A^{\mu}$ is the analogous axial current (with $\gamma^\mu$ replaced by $\gamma^5 \gamma^\mu$). The sum is performed over all SM fermion species  $X$. For convenience, we wrote separately the terms containing $N$ which plays the role of DM  and can be Dirac or right-handed Majorana fermion. The interaction \eqref{Lint} vanishes only if $\alpha=0$, $\beta=\pm 1$, and in what follows we do not consider this particular choice of the couplings. 

Adding the Holst term to Eq.~(\ref{S_f}) modifies the couplings in Eq.~(\ref{Lint}). Furthermore, including a scalar field coupled non-minimally to gravity results in additional scalar-fermion interactions. Below we focus on the four-fermion interaction (\ref{Lint}) whose contribution to the DM production dominates for small masses of $N$. We discuss the general case in appendix A, which includes Refs.~\cite{Immirzi:1996dr,Immirzi:1996di}.
\\

\emph{Thermal production of singlet fermions.}---The four-fermion interaction (\ref{Lint}) opens up the production channel of $N$-particles through the annihilation of the SM fermions $X$, via the reaction $X + \bar{X} \to N + \bar{N}$.\footnote{The production of singlet fermions due to some higher-dimensional operators was considered in \cite{Bezrukov:2011sz}. However, the four-fermion interaction which appears in EC gravity was not accounted for.} The kinetic equation corresponding to this reaction takes the form
\begin{equation}
    \left(\frac{\partial}{\partial t}-H q_{i} \frac{\partial}{\partial q_{i}}\right) f_N(t, \vec{q})=R(\vec{q}, T) \;,
    \label{kin_eq}
\end{equation}
where $f_N$ is the phase-space density of $N$, $H$ is the Hubble rate and $R$ is the collision integral, also referred to as a production rate. In an isotropic background, both $f_N$ and $R$ depend only on the absolute value of the spatial momentum $|q|\equiv |\vec{q}|$.

In what follows, we assume that all SM particles, including fermions, are in thermal equilibrium at the moment of DM production. To check the validity of this assumption, one would need a careful examination of dynamics of bosonic and fermionic SM species at and after preheating, which goes beyond the scope of the present paper. We expect, however, that deviations from thermality do not change qualitatively our results.

As long as the concentration of $N$ remains small and we can neglect the inverse processes, the collision integral in \cref{kin_eq} reads 
\begin{equation}
R=\frac{1}{2 |q|}  \sum_X \int \diff\Pi\: |\overline{\mathcal{M}}_X|^2 \,f_X( p_1) f_{\bar{X}}(p_2) \;,
\label{rate}
\end{equation}
where the sum runs over the SM species ($24$ left- and $21$ right-handed fermions), $\overline{\mathcal{M}}_X$ is the amplitude of the process summed over all spinor indices, and $f_X$ is the distribution of $X$ which we assume to be the Fermi-Dirac one. The differential $\diff\Pi$ is the phase space volume element accounting for energy conservation; see appendix B for its definition, where also details on the derivations of subsequent \crefrange{res_rate}{general_result} are given. 
The typical values of the momenta in \cref{rate} are large compared to the mass of $N$, so we neglect all masses when computing $\overline{\mathcal{M}}_X$.
Introducing the dimensionless variable $y = E/T$, where $T$ is the temperature of the cosmic plasma   and $E = |q|$,
we arrive at 
\begin{equation}
     R(E,T) = T^5 \, \frac{C_f(\alpha, \beta)}{M_P^4} r(y) \;.
     \label{res_rate}
 \end{equation} 
Here $C_f(\alpha, \beta)$ is a combination of the non-minimal fermion couplings whose precise form depends on whether $N$ is Majorana ($f=M$) or Dirac ($f=D$): 
\begin{equation}
     C_M = \dfrac{9}{4}\left(24a^4 + 21 b_+^4 \right) \:, ~ C_D = \dfrac{9}{4}\left(45a^4+ 21 b_+^4 + 24 b_-^4\right) \:,
 \label{Coeff}
\end{equation}
where $a^2=1+\alpha^2-\beta^2$, $b_{\pm}^2=1-(\alpha\pm\beta)^2$. Next, $r(y)$ is a function that we compute numerically (following~\cite{Asaka:2006nq}). It is accurately approximated by\footnote{This expression is exact if instead of the Fermi-Dirac distribution, one uses the Boltzmann distribution for $f_X$.}
\begin{equation}
   r(y) \simeq \frac{1}{24\pi^3} y f_X \;.
   \label{r_approx}
\end{equation}
\cref{kin_eq} can now be easily integrated, leading to
 \begin{equation}
     f_N(y) = \frac{C_f \, T^3_{\mathrm{prod}} \,M_0(T_{\mathrm{prod}})}{3 \, M_P^4} r(y) \;,
     \label{fN}
 \end{equation}
where $T_{\mathrm{prod}}$ is the temperature at which the DM production begins,
$M_0(T)=M_P \sqrt{\frac{90}{\pi^2 g_{\mathrm{eff}}(T)}}$, and ${g_{\mathrm{eff}}(T_{\rm prod})=106.75}$ is the number of effectively massless degrees of freedom at high temperature. 
Plugging in the numbers, we obtain for the abundance of $N$-particles:\footnote{In deriving this result, we assume that all other possible interactions of $N$-particles with particles of the SM are not essential for the DM production.}
\begin{equation}
    \frac{\Omega_N}{\Omega_{DM}}\simeq 3.6 \cdot 10^{-2} \, C_f \left( \frac{M_N}{10~\text{keV}}\right) \, \left( \frac{T_\mathrm{prod}}{M_P} \right)^3 \;,
    \label{general_result}
\end{equation}
where $\Omega_{DM}$ is the observed DM abundance and the coefficient $C_f$ is defined in \cref{Coeff}. \cref{general_result} shows that, depending on the value of $C_f$, the right amount of DM can be generated in a broad range of fermion masses $M_N$. 

To proceed further, we need an estimate for the production temperature. We obtain it within the framework of Higgs inflation in the Palatini formulation of gravity~\cite{Bauer:2008zj,Tenkanen:2020dge}. In this model, preheating is almost instantaneous \cite{Rubio:2019ypq}, and one can take $T_{\rm prod}\sim T_{\rm reh}$ where
\begin{equation}
T_{\rm reh}\simeq \left(\frac{15\lambda}{2\pi^2 g_{\rm eff}}\right)^{\frac{1}{4}}\frac{M_P}{\sqrt{\xi}}
\end{equation}
is the preheating temperature, $\lambda$ is the Higgs field self-coupling and $\xi$ is the non-minimal coupling of the Higgs field to the Ricci scalar. Both $\lambda$ and $\xi$ are taken at a high energy scale. Using $\xi=10^7$ and $\lambda=10^{-3}$ \cite{Shaposhnikov:2020fdv}, we obtain $T_{\rm prod} \simeq 4\times 10^{13}$~GeV.

Now we can investigate two particularly interesting cases. The first one is the limit of vanishing non-minimal couplings, $\alpha = \beta = 0$. Then, from \cref{general_result} we obtain that $\Omega_N\simeq\Omega_{DM}$ if $M_N\simeq 6\times 10^8$~GeV for the Majorana fermion and $M_N\simeq 3\times 10^8$~GeV for the Dirac fermion. We conclude that heavy fermion DM can be produced in EC gravity even if the action of the EC theory is identical to that of the metric theory.\footnote{Such a heavy DM can also be produced due to graviton exchange between fermionic currents.} Interestingly, the given bounds on $M_N$ are close to the bound $M_N\lesssim 10^9$~GeV above which $N$-particles are overproduced due to the varying geometry at the radiation-dominated stage of the Universe \cite{Mamaev:1976zb, Kuzmin:1998uv,Chung:1998ua}.

The second case corresponds to setting $\alpha \sim \beta \sim \sqrt{\xi}$. With this choice, the scale of suppression of the interaction (\ref{Lint}) coincides with the inflationary cutoff scale which in Palatini Higgs inflation is of the order of $M_P/\sqrt{\xi}$ \cite{Bauer:2010jg}. 
For both the Majorana and Dirac cases, \cref{general_result} becomes
\begin{align}
  \scalebox{0.92}[1]{$\dfrac{\Omega_N}{\Omega_{DM}}\simeq 1.4  \, \dfrac{\sqrt{\xi} \lambda^{3/4}}{g_{\rm eff}^{3/4}}\, \dfrac{\left(\alpha + \beta\right)^4}{\xi^2} \left( \dfrac{M_N}{10~\text{keV}}\right) \,\left( \dfrac{T_\mathrm{prod}}{T_{\rm reh}} \right)^3 $} \;. 
    \label{Palatini_result}  
\end{align}
Thus, the right amount of DM is generated for a keV-scale $M_N$.\\

\emph{Einstein-Cartan portal to warm dark matter.}---Let us discuss in more detail the second choice of the non-minimal couplings. Then, $N$ is an example of warm DM, and its free-streaming length affects structure formation. Since DM is produced at very high temperatures, its spectrum is redshifted. Consequently, the average momentum is only $\simeq 0.61$ of the equilibrium momentum at $T=1~$MeV. Depending on the history of reionization, such colder DM candidate can provide a better fit to the Lyman-$\alpha$ data than pure cold DM \cite{Garzilli:2018jqh,Garzilli:2019qki}.
\begin{figure}[t]
    \centering
    \includegraphics[width=0.47\textwidth]{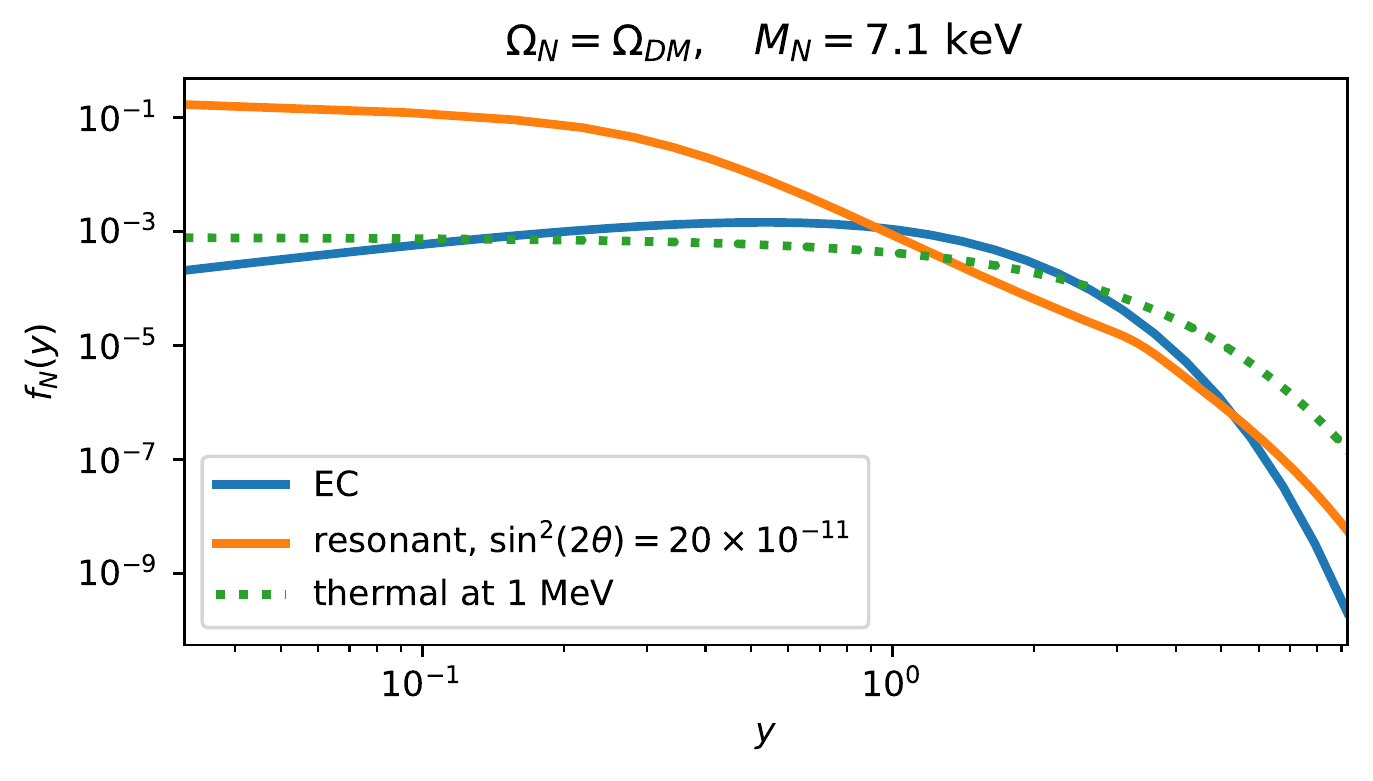}
    \caption{The DM distribution function at $T=1$~MeV. The normalisation is such that $\Omega_N=\Omega_{DM}$ for the benchmark mass $M_N=7.1$~keV used in \cite{Ghiglieri:2015jua}. The blue curve shows the red-shifted spectrum (\ref{fN}) whereas the orange one shows the spectrum of resonantly produced sterile neutrino~\cite{Ghiglieri:2015jua}. For comparison, we also show the thermal spectrum (the green dotted line).}
    \label{fig:spectrum}
\end{figure}

In \cref{fig:spectrum} we show the spectrum $f_N$ of DM produced in EC gravity. For comparison, we also show the spectrum of resonantly produced sterile neutrino taken from \cite{Ghiglieri:2015jua}\footnote{This result depends on the assumptions about the lepton asymmetry and the values of Yukawa coupling. The curve in \cref{fig:spectrum} corresponds to the ``case a'', $\sin ^{2}(2 \theta)=20 \times 10^{-11}$ in terms of \cite{Ghiglieri:2015jua}.} and the thermal spectrum at $1~$MeV. We see that the DM momentum distribution (\ref{fN}) has a unique form and  differs from both non-resonantly~\cite{Dodelson:1993je,Dolgov:2000ew} and resonantly produced sterile neutrinos~\cite{Shi:1998km,Laine:2008pg,Abazajian:2001nj,Venumadhav:2015pla,Ghiglieri:2015jua}. This potentially allows one to distinguish DM produced via the EC portal from other DM candidates.
We conclude that, at least for certain values of $M_N$, the distribution of DM in the Universe can bear the information about the gravity-induced fermionic interactions.\\

\emph{$\nu$MSM.}---The lack of a DM candidate is a famous shortcoming of the SM, but it is not the only one. It also cannot explain neutrino oscillations and the baryon asymmetry of the Universe. A possible way to address these three issues at once is the Neutrino Minimal Standard Model ($\nu$MSM), which extends the particle content of the SM by three right-handed neutrinos $N_{1,2,3}$~\cite{Asaka:2005an,Asaka:2005pn}. One of them, $N_1$, can be the DM candidate. A lower bound on its mass, $M_{N_1}\gtrsim 1~$keV, comes from small-scale structures in the matter power spectrum, as inferred, e.g., from Lyman-$\alpha$ measurements. The mixing of $N_1$ with active neutrinos is bounded from above by X-ray constraints. The other two right-handed neutrinos $N_{2,3}$ have nearly-degenerate GeV-scale masses and are responsible for generating the baryon asymmetry. Moreover, the parameters of the model can be chosen such that the observed pattern of neutrino oscillations is obtained
(for reviews see \cite{Boyarsky:2009ix,Boyarsky:2018tvu}).

So far, the production of DM sterile neutrino was associated with the mixing of $N_1$ with ordinary neutrinos and typical temperatures $100-300$ MeV \cite{Dodelson:1993je,Shi:1998km}. The so-called non-resonant production of $N_1$ \cite{Dodelson:1993je} has already been excluded by X-ray searches of radiatively decaying DM \cite{Boyarsky:2018tvu}. The resonant production of DM sterile neutrino \cite{Shi:1998km} requires large lepton asymmetries which can be produced in interactions of $N_{2,3}$ \cite{Shaposhnikov:2008pf}. It can be successful albeit a fine-tuning of parameters is required \cite{Canetti:2012vf,Canetti:2012kh,Ghiglieri:2020ulj}. For resonantly produced DM, the non-observation of X-ray decays of DM in galactic halos leads to an upper bound on the mass, ${M_{N_1} \lesssim 50}$~keV \cite{Boyarsky:2018tvu}, thereby constraining $M_{N_1}$ in a quite narrow range.

The present work supplies the $\nu$MSM with a different mechanism to produce the $N_1$-particles. Sterile neutrino DM produced this way may be absolutely stable and, therefore, is not subject to any X-ray constraints. This opens up a new interesting mass window up to ${M_{N_1}\sim 10^8}$~GeV. The prediction of the $\nu$MSM that the lightest active neutrino is effectively massless remains in force also for large values of $M_{N_1}$; see appendix C.\footnote{Note that this prediction can potentially be tested by the Euclid space mission~\cite{Audren:2012vy}.}

It is very intriguing that if the perturbative cutoff is universal for both the fermion and scalar-gravity sectors of the EC theory ($\sim M_P/\sqrt{\xi}$), then the mass of the DM sterile neutrino is required to be in the keV range. This is the domain where the warm nature of the DM particle is most visible and in which
the most intensive searches of the radiatively decaying DM are being carried out.

We note that Majorana fermions $N_{2,3}$  will be also produced by the same mechanism. Their number density is the same as that of $N_1$, $n_{2,3}\simeq  n_1$, and the latter is fixed by the DM abundance $n_1\simeq  10^{-2}n_{eq} (10 \:\rm{keV}/M_{N_1})$. However, this density is too small to affect the analysis of the $\nu$MSM leptogenesis carried out previously (e.g.~\cite{Klaric:2020lov}), see appendix D.\\ 

\emph{Discussion and outlook.}---Once gravity is coupled to matter, such as fermions or a non-minimally coupled scalar field, its different formulations are no longer equivalent. As long as they are consistent, only observations can help us to distinguish between them. In the present work, we have shown that properties of fermionic dark matter may be able to discern the Einstein-Cartan (EC) theory of gravity from the most commonly used metric formulation. In particular, the universal dimension-six interactions of fermionic currents in EC theory can cause the production of the observed amount of dark matter for a wide range of fermion masses. Moreover, they lead to a characteristic momentum distribution of dark matter, which can serve to confirm or exclude our proposed production mechanism. 

On the one hand, these findings are relevant for any extension of the Standard Model (SM) by sufficiently heavy fermions. On the other hand, an exciting unified picture of gravity and the SM emerges. It relies on the EC formulation of gravity and the extension of SM by three right-handed neutrinos, i.e. the Neutrino Minimal Standard Model ($\nu$MSM) \cite{Asaka:2005an,Asaka:2005pn}. When non-minimally coupled to the Ricci scalar, the Higgs field can assume the role of the inflaton \cite{Bezrukov:2007ep}. As discussed in \cite{Shaposhnikov:2020geh}, the resulting inflationary scenario generalizes the model of Palatini Higgs inflation \cite{Bauer:2008zj} and is fully compatible with observations. On its own, the $\nu$MSM is able to explain neutrino oscillations, baryogenesis and provides a dark matter candidate in the form of a right-handed neutrino. The result of the present work is that the EC formulation of gravity can lead to the production of this sterile neutrino in an amount that matches the observed abundance of dark matter. Finally, we remark that the Palatini formulation of gravity and, by generalization, EC gravity is convenient for addressing the question of the big difference between the Electroweak and the Planck scales \cite{Shaposhnikov:2020geh} (see also \cite{Shaposhnikov:2018xkv, Shaposhnikov:2018jag, Shkerin:2019mmu, Karananas:2020qkp}).

\begin{acknowledgments}
\emph{Acknowledgments.}---We thank Alexey Boyarsky and Oleg Ruchayskiy for useful discussions. The work was supported by ERC-AdG-2015 grant 694896 and by the Swiss National Science
Foundation Excellence grant 200020B\underline{ }182864.
\end{acknowledgments}


\vspace{1em}
\appendix

\emph{Appendix A.}---\cref{S_f} is not the most general dimension-four Lagrangian allowed in EC theory. Due to the presence of torsion, its gravitational part can be extended by adding the Holst term \cite{Hojman:1980kv, Nelson:1980ph, Castellani:1991et, Holst:1995pc}:
\begin{equation}
\Delta\mathcal{L}_{\rm grav.}=\frac{M_P^2}{4\bar{\gamma}}\epsilon^{\mu\nu\rho\sigma}R_{\mu\nu\rho\sigma} \;.
\end{equation}
The coupling $\bar{\gamma}$ is called the Barbero-Immirzi parameter \cite{Immirzi:1996dr,Immirzi:1996di}. The presence of the Holst term modifies the coefficients in the four-fermion interaction terms: \cref{Lint} becomes \cite{Shaposhnikov:2020frq}
\begin{equation}
\begin{split}\label{4f-Holst}
\mathcal{L}_{4f} = &\frac{3}{16M_P^2} \frac{\alpha^{2} \bar{\gamma}^{2}}{\bar{\gamma}^{2} +1}V^\mu V_\mu 
\\
&- \frac{3}{8M_P^2} \frac{\alpha \bar{\gamma}^{2} }{\bar{\gamma}^{2} +1}\left(\frac{1}{\bar{\gamma} }-\beta\right) V^\mu A_\mu 
\\
&- \frac{3}{16M_P^2} \frac{\bar{\gamma}^{2} }{\bar{\gamma}^{2} +1}\left(1+\frac{2 \beta}{\bar{\gamma} }-\beta^{2}\right)A^\mu A_\mu \;.
\end{split}
\end{equation}
In the limit ${\bar{\gamma}\to\infty}$, \cref{Lint} is restored. In the limit ${\bar{\gamma}\to 0}$, all coefficients vanish. This agrees with the fact that the latter limit corresponds to the case of vanishing torsion (see, e.g., \cite{hep-th/0507253}). 

The action of EC theory can also be extended by including non-minimal coupling of a scalar field $h$ (such as the Higgs field) to gravity. This coupling does not contribute to the four-fermion interaction in the metric description. Instead, it gives rise to scalar-fermion interaction terms~\cite{Shaposhnikov:2020frq}.
The scalar-fermion part of the action reads
\begin{equation}
\begin{aligned}
    S_{s f} &= \int \diff^4x\sqrt{-g}\frac{3\alpha}{4}\left( \frac{\partial_\mu\Omega^2}{\Omega^2}+\frac{\gamma}{\gamma^2+1}\left( \frac{\partial_\mu\bar{\eta}}{\Omega^2}+\partial_\mu\gamma \right) \right) V^\mu \\
& + \int \diff^4x\sqrt{-g}\frac{3}{4}\left( \beta\frac{\partial_\mu\Omega^2}{\Omega^2}+\frac{1+\gamma\beta}{\gamma^2+1}\left( \frac{\partial_\mu\bar{\eta}}{\Omega^2}+\partial_\mu\gamma \right) \right) A^\mu 
    \end{aligned}
\end{equation}
with $\Omega^{2}=1+\frac{\xi h^{2}}{M_{P}^{2}}$ and
\begin{equation}
    \gamma(h) =\frac{1}{\bar{\gamma}\Omega^2}\left(1+\frac{\xi_{\gamma} h^2}{M_P^2}\right) \,, ~~~
    \eta(h) =\frac{\xi_\eta h^2}{M_P^2} \,.
\end{equation}
In the expressions above $h$ is the Higgs field in the unitary gauge, whereas $\xi_\gamma$ and $\xi_\eta$ are the non-minimal couplings of $h$ to the Holst term and Nieh-Yan invariant respectively, see~\cite{Shaposhnikov:2020frq} for details.
We study this action in the limit $h\ll M_P/\sqrt{\xi}$ and $h\ll M_P/\sqrt{\xi_\gamma}$.
Upon integrating by parts, we obtain the following interaction Lagrangian:
\begin{equation}
\begin{aligned}
    \mathcal{L}_{s f} & = -\frac{3\alpha}{4}\left(\xi+\frac{\bar{\gamma}\xi_\eta+\xi_\gamma-\xi}{\bar{\gamma}^2+1}\right)\left(\frac{h}{M_P}\right)^2\partial_\mu V^\mu \\
    & - \frac{3}{4}\left(\beta\xi+\frac{(\bar{\gamma}+\beta)(\bar{\gamma}\xi_\eta+\xi_\gamma-\xi)}{\bar{\gamma}^2+1}\right)\left(\frac{h}{M_P}\right)^2\partial_\mu A^\mu \;.
\end{aligned}
\end{equation}
The vector current is conserved, $\partial^\mu V_\mu = 0$, whereas the divergence of the axial current is proportional to the fermion mass $M_N$. As a result, in the presence of the additional terms non-minimally coupled to the Higgs field, a new channel $h + h \to N + \bar{N}$ contributes to the production of singlet fermions. The amount of DM produced via this channel can be calculated analogously to the four-fermion case. The abundance of $N$ generated via the Higgs annihilation is given by
\begin{equation}
\begin{aligned}
    \dfrac{\Omega_N}{\Omega_{DM}}\simeq & 3.2 \,
   \left(\beta\xi + \frac{(\bar{\gamma}+\beta)(\bar{\gamma}\xi_\eta+\xi_\gamma-\xi)}{\bar{\gamma}^2+1} \right)^2\\
    & \times \, \left( \dfrac{T_{\rm prod}}{M_P}\right)^3 \left( \dfrac{M_N}{T_{\rm prod}}\right)^2 \left( \dfrac{M_N}{10~\rm keV}\right) \;,
    \label{Omega_Higgs}
     \end{aligned}
\end{equation}
where we specialized to a Dirac fermion $N$; the Majorana case would lead to an additional factor of $1/2$. 

We see that, as compared to the four-fermion channel \cref{general_result}, the DM production due to the scalar-fermion interaction is suppressed by a factor of $(M_N/T_{\rm prod})^2$. This means that for small masses of the DM particle, the four-fermion channel provides the dominant contribution to the DM abundance. Moreover, it is possible to choose the non-minimal couplings such that the scalar-fermion channel vanishes regardless the mass of $N$. For example, this is the case for
\begin{equation}
\frac{1}{\bar{\gamma}}=\xi_\eta=\beta=0 \;,
\end{equation}
or when the coupling $\beta$ is given by the following combination of other parameters:
\begin{equation}
\beta=\frac{-\xi_\eta\bar{\gamma}^2+\xi\bar{\gamma}-\xi_\gamma\bar{\gamma}}{\xi\bar{\gamma}^2+\xi_\eta\bar{\gamma}+\xi_\gamma} \;.
\end{equation}
In a more general situation, the observed amount of DM can be generated mainly in the scalar-fermion channel in certain regions of the parameter space. As an example, let us take $\bar{\gamma} =1$, $\xi \simeq 10^{7}$, $\xi_{\gamma}=\xi_\eta=\alpha=0$ corresponding to a particular realization of Holst inflation \cite{Shaposhnikov:2020geh}. If we then take $\beta = \sqrt{\xi}$, the contribution of the process $h + h \to N + \bar{N}$ is suppressed compared to the four-fermion one by several orders of magnitudes. If, on the other hand, we take $\beta = 0$, the scalar-fermion channel dominates over the four-fermion one, and the observed amount of DM is generated for $m_N=3.6 \times 10^7~\text{GeV}$.

We note that the value of $\Omega_N$ depends on $T_{\rm prod}$, which in turn is sensitive to the details of preheating. These may generically depend on the non-minimal couplings. Nevertheless, the main conclusions about the possibility of gravitational production of DM via the four-fermion or scalar-fermion interactions and about the relative strength of different production channels are independent of the details of preheating.
\\
    
\begin{figure}[h]
    \centering
    \includegraphics[width=0.47\textwidth]{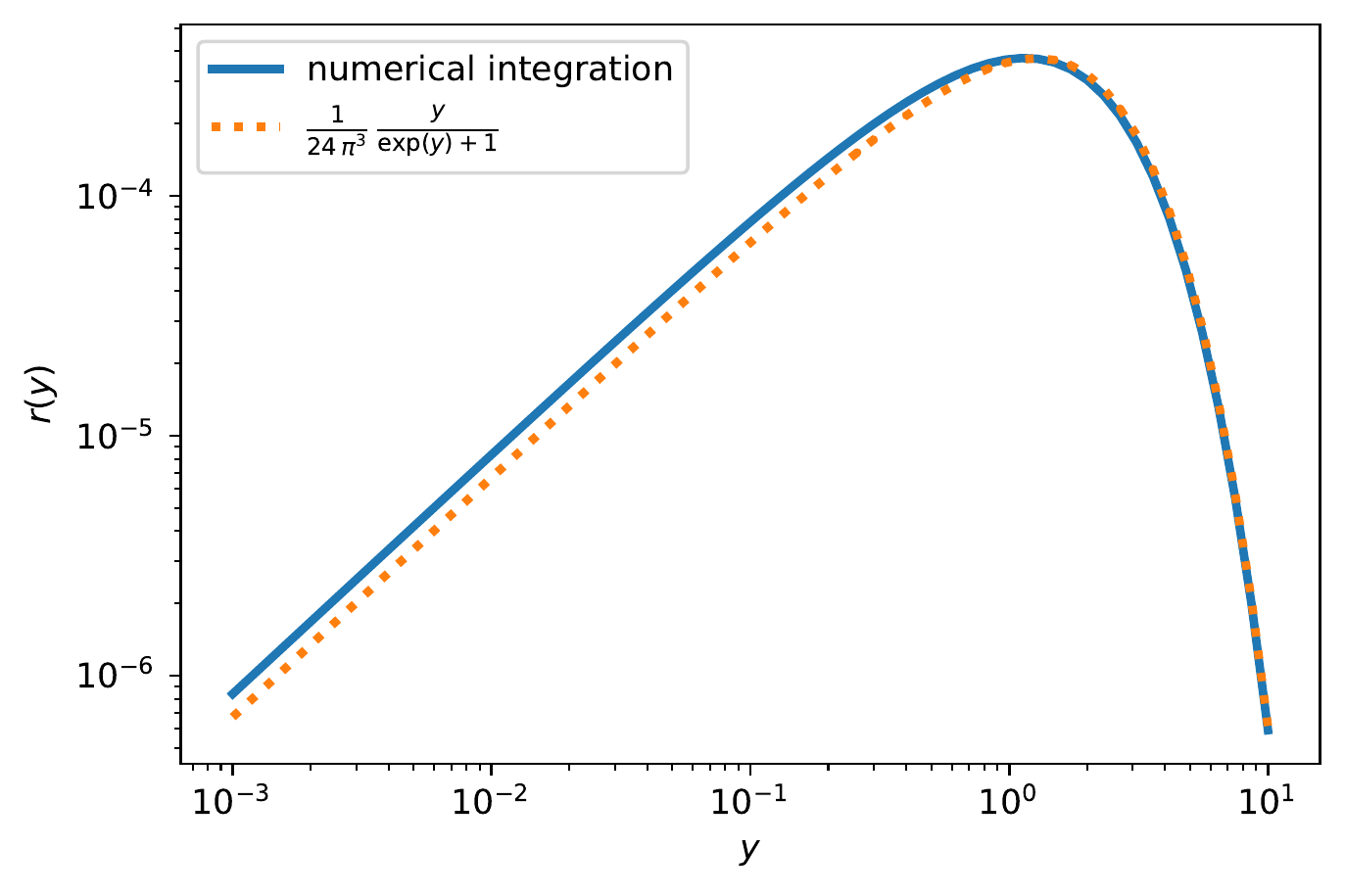}
    \caption{The dimensionless production rate $r(y)$, $y=E/T$. The DM spectrum $f_N(y)$ is proportional to $r(y)$. The blue solid line shows the result of the numerical integration, whereas the orange dotted line shows the analytic approximation.}
    \label{fig:rate}
\end{figure}

\emph{Appendix B.}---Here we provide some details about the calculation of the DM abundance.
We start from computing the amplitude entering \cref{rate}.
It is convenient to split the production channels as
\begin{align}
X_L(p_1) + \bar{X}_L(p_2) & \to N(q) + \bar{N}(p_3)\;,\label{prodL}\\
X_R(p_1) + \bar{X}_R(p_2) &\to N(q) + \bar{N}(p_3)\;.\label{prodR}
\end{align}
As mentioned in the main text, we consider both Majorana and Dirac masses for $N$. In the Majorana case, we assume that $N$ is a right-handed fermion, and then the amplitudes corresponding to \cref{prodL} and \cref{prodR} read
\begin{align}
|\overline{\mathcal{M}}_{X_L}|^2 &= \frac{64}{M_P^4}\left[ C_{A A}-C_{V V} \right]^{2} \left( p_1 \cdot p_3 \right)  \left( p_2 \cdot q \right)\;,\label{M_ML}\\
|\overline{\mathcal{M}}_{X_R}|^2 &= \frac{64}{M_P^4}\left[ C_{A A}+C_{V A} + C_{V V} \right]^{2} \left( p_1 \cdot q \right)  \left( p_2 \cdot p_3 \right)\;,\label{M_MR}
\end{align}
where by $C_{VV}$, $C_{VA}$ and $C_{AA}$ we denote the coefficients of the vector-vector, vector-axial and axial-axial currents, respectively. In the Dirac case, the amplitudes are
\begin{align}
|\overline{\mathcal{M}}_{X_L}|^2 &= \frac{64}{M_P^4} \left\{ \left[ C_{A A}-C_{V V} \right]^{2} \left( p_1 \cdot p_3 \right)  \left( p_2 \cdot q \right) \right. \nonumber\\
&+\left. \left[ C_{A A}-C_{V A} + C_{V V} \right]^{2} \left( p_1 \cdot q \right)  \left( p_2 \cdot p_3 \right) \right\}\;, 
\label{M_DL} \nonumber \\
|\overline{\mathcal{M}}_{X_R}|^2 &= \frac{64}{M_P^4} \left\{ \left[ C_{A A}-C_{V V} \right]^{2} \left( p_1 \cdot p_3 \right)  \left( p_2 \cdot q \right) \right.\\
&+\left. \left[ C_{A A}+C_{V A} + C_{V V} \right]^{2} \left( p_1 \cdot q \right)  \left( p_2 \cdot p_3 \right) \right\}\;.
\label{M_DR}
\end{align}
If there is no asymmetry between particles ($X$) and anti-particles ($\bar{X}$), the integral in \cref{rate} is symmetric under the replacement $p_1 \leftrightarrow p_2$. As a consequence, both structures yield the same function $r(y)$ (cf. \cref{res_rate}).
The dimensionless function is defined simply as $r(y) = r(y T, T) / T^5$, where $y = E/T$ and 
\begin{equation}
\begin{aligned}
&r(E,T) =  \,\frac{1}{2 E}  \sum_X \int \frac{\Diff3 \vec{p}_1}{(2\pi)^3 \, 2 E_1}
\frac{\Diff3 \vec{p}_2}{(2\pi)^3 \, 2 E_2} \frac{\Diff3 \vec{p}_3}{(2\pi)^3 \, 2 E_3} \\
& \times (2\pi)^4\delta^{(4)}\left( p_1+p_2-q-p_3 \right) \, \left( p_1 \cdot q \right)  \left( p_2 \cdot p_3 \right) \\
& \times f_X( p_1) f_{\bar{X}}(p_2) \;.
\end{aligned}
\label{def_r}
\end{equation}
The integration in \cref{def_r} can be performed numerically following the recipe in~\cite{Asaka:2006nq}.
We assume that all SM fermions thermalize rapidly after inflation and, hence, we take the Fermi-Dirac distribution $f_X = f_{\bar{X}} = (\exp(y)+1)^{-1}$.
Note that the computation of the exact fermion spectrum right after inflation is a highly non-trivial task, and we do not attempt to address it here.
The result of the numerical integration is shown in \cref{fig:rate}. In the same figure we also show the analytic approximation given in \cref{r_approx}.

We can rewrite \cref{kin_eq} in the form
\begin{equation}
T\, H \,\frac{\partial f_{N}}{\partial T}=-\frac{C_f}{M_{P l}^{4}}T^5\, r(y)\;,
\label{kin_eq2}
\end{equation}
where the coefficient $C_f$ ($f = M,\, D$) is given by \cref{Coeff}. In the more general case of non-vanishing $1/\bar{\gamma}$ the coefficient can be read from \cref{4f-Holst} and \crefrange{M_ML}{M_DR}. Now \cref{kin_eq2} can be easily integrated leading to \cref{fN}.

In order to compute the DM abundance, it is convenient to introduce the variable $Y_N=n_N/s$ which remains constant in the expanding Universe. Here $s$ corresponds to the entropy density and $n_N=2\cdot 1/(2\pi)^3 \int \diff^3 \vec{q}\ f_N$ is the number density. In the latter, the factor $2$ accounts for $\bar{N}$. Then we get
\begin{equation}
Y_N = \frac{1}{\frac{2\pi^2}{45} g_{\mathrm{eff}} T^3} \, 2\cdot T^3 \int_0^\infty \frac{\diff y \,y^2}{2 \pi^2} \,f_N(y)\;.
\end{equation}
The abundance of $N$ is given by
\begin{equation}
\Omega_{N}\equiv \frac{M_{N} n_{N}}{\rho_{cr}} = \frac{M_N Y_N}{\rho_{cr}/s}\;,
\label{Omega_N}
\end{equation}
where $\rho_{cr}$ is the critical energy density of the Universe. Plugging in the numbers, we arrive at \cref{general_result}.\\

\emph{Appendix C.}---As was mentioned in the main text, the lightest active neutrino is practically massless in the $\nu$MSM regardless the mass of the DM sterile neutrino $N_1$.\footnote{The statement about keV-scale DM has been made in~\cite{Asaka:2005an}.} Indeed, in the seesaw mechanism the contribution of $N_1$ to the mass matrix $m_\nu$ of the active neutrinos can be estimated as
\begin{equation}
\delta m_\nu \sim M_{N_1} \theta^2, \quad \theta^2\equiv \sum_\alpha |\theta_\alpha|^2,
\label{m_contrib}
\end{equation}
where $\theta_\alpha$ is the mixing between $N_1$ and the neutrino of flavor $\alpha = e, \mu, \tau$.
The mixing $\theta^2$ implies that $N_1$ is not absolutely stable and processes like $N_1 \to \nu \nu \nu$, $N_1 \to \nu \gamma$ are allowed. The stability of DM and the non-observation of a monochromatic signal from the decay $N_1\to \nu \gamma$ constrain $\theta^2$ and, by virtue of \cref{m_contrib}, the contribution of $N_1$ to the active neutrino mass matrix. 
The other two right-handed neutrinos $N_{2,\,3}$ of the $\nu$MSM can provide two contributions to $m_\nu$ matching the two observed mass differences (the solar and the atmospheric ones). Therefore, the contribution of $N_1$ can be identified with the mass of the lightest active neutrino.

In the mass range $1-50$~keV where the resonant production is effective, the combination $\theta^2$ has been bounded from above by the X-ray observations~\cite{Boyarsky:2006jm} (for a recent review, see~\cite{Boyarsky:2018tvu}). One can expect that for larger masses of $N_1$, the X-ray (and gamma) constraints also lead to the strongest bounds. However, for our purposes the requirement of the DM stability is already enough. Indeed, the width of the three-body decay of $N_1$ reads \cite{Pal:1981rm,Barger:1995ty}
\begin{equation}
\Gamma_{N_1 \rightarrow 3 \nu}=\frac{G_{F}^{2} M_{N_1}^{5}}{96 \pi^{3}} \theta^2,
\end{equation}
where $G_F$ is the Fermi constant. Requiring that $1/\Gamma_{N_1 \rightarrow 3 \nu}$ exceeds the age of the Universe, we arrive at an upper bound on $\theta^2$. Using \cref{m_contrib} we get 
\begin{equation}
\delta m_\nu \lesssim 3.3 \text{~eV}\, \times \left( \frac{10\text{~keV}}{M_{N_1}} \right)^4. 
\end{equation}
We see that already for $M_{N_1}=100$~keV, the requirement of the stability of $N_1$ yields $\delta m_\nu $ being $\sim 25$ times less than the solar mass difference $\sqrt{\Delta m_{sol}}\simeq 8.6 \times 10^{-3}$~eV. It is also clear that this bound rapidly gets stronger when $M_{N_1}$ increases.  
\\

\emph{Appendix D.}---The possibility of successful leptogenesis with two Majorana fermions $N_{2,3}$ has first been shown in Ref.~\cite{Asaka:2005pn}. Inspecting the perturbative solution of Ref.~\cite{Asaka:2005pn} one can see that the produced Baryon Asymmetry of the Universe (BAU)
is proportional to $( n_{2,3}-n_{eq})/s$, where $n_{2,3}$ is the number density of  Majorana fermions $N_{2,3}$ and $n_{eq}$ is the equilibrium number density of relativistic fermions, and $s$ is the entropy density. The fermions $N_{2,3}$ are produced by the same processes as $N_1$. Since the masses can be safely neglected, the yields $Y_X=n_X/s$ will be the same for all right-handed neutrinos. Using \cref{Omega_N} and assuming that $N_1$ forms all DM, we can find that
\begin{equation}
   Y_{N_{2,3}} =  Y_{N_1} \simeq 4.4\times 10^{-5} \frac{10~\text{keV}}{M_{N_1}}\,.
\end{equation}
This value should be compared with the equilibrium yield $Y_{eq}=n_{eq}/s\simeq 1.9\times 10^{-3}$. As one can see, for a $10$~keV $N_1$, the baryogenesis results are affected at a percent level. 
We have also verified numerically (using the kinetic equations of ref.~\cite{Klaric:2020lov}), that the same conclusion holds in the regions of the parameter space where the perturbative solutions is not applicable.

\bibliography{Refs}

\end{document}